# Attractive forces in microporous carbon electrodes for capacitive deionization

P.M. Biesheuvel,[1,2,3] S. Porada,[1] M. Levi,[4] and M.Z. Bazant[5,6]

[1] *Wetsus, centre of excellence for sustainable water technology, Agora 1, 8934 CJ Leeuwarden, The Netherlands.* [2] *Laboratory of Physical Chemistry and Colloid Science, Wageningen University, Dreijenplein 6, 6703 HB Wageningen, The Netherlands.* [3] *Department of Environmental Technology, Wageningen University, Bornse Weilanden 9, 6708 WG Wageningen, The Netherlands.* [4] *Department of Chemistry, Bar-Ilan University, Ramat-Gan 52900, Israel.* [5] *Department of Chemical Engineering, Massachusetts Institute of Technology, Cambridge, MA 02139, USA.* [6] *Department of Mathematics, Massachusetts Institute of Technology, Cambridge, MA 02139, USA.*

**Abstract**

The recently developed modified Donnan (mD) model provides a simple and useful description of the electrical double layer in microporous carbon electrodes, suitable for incorporation in porous electrode theory. By postulating an attractive excess chemical potential for each ion in the micropores that is inversely proportional to the total ion concentration, we show that experimental data for capacitive deionization (CDI) can be accurately predicted over a wide range of applied voltages and salt concentrations. Since the ion spacing and Bjerrum length are each comparable to the micropore size (few nm), we postulate that the attraction results from fluctuating bare Coulomb interactions between individual ions and the metallic pore surfaces (image forces) that are not captured by mean-field theories, such as the Poisson-Boltzmann-Stern model or its mathematical limit for overlapping double layers, the Donnan model. Using reasonable estimates of the micropore permittivity and mean size (and no other fitting parameters), we propose a simple theory that predicts the attractive chemical potential inferred from experiments. As additional evidence for attractive forces, we present data for salt adsorption in uncharged microporous carbons, also predicted by the theory.

## 1. Introduction

Electrodes made of porous carbons can be utilized to desalinate water in a technique called capacitive deionization (CDI) in which a cell is constructed by placing two porous carbon electrodes parallel to one another [1-12]. A cell voltage difference is applied between the electrodes, leading to an electrical and ionic current in the direction from one electrode to the other. The water flowing through the cell is partially desalinated because ions are adsorbed in their respective counterelectrode. CDI is in essence a purely capacitive process, based on the storage (electrosorption) of ions in the electrical double layer (EDL) that forms within the electrolyte-filled micropores of the carbon upon applying a cell voltage (the measurable voltage difference applied between anode and cathode). In CDI, the cathode (anode) is defined as the electrode that adsorbs the cations (anions) during the desalination step. Note that while counterions are adsorbed, co-ions are expelled from the EDLs, leading to a diminished desalination and a so-called "charge efficiency" $\Lambda$ lower than unity [13-20]. The charge efficiency $\Lambda$ is defined for a 1:1 salt solution (NaCl) as the ratio of salt adsorption by a CDI electrode cell pair, divided by the charge stored in an electrode. It is typically defined for a cycle where the salt adsorption step is long enough for equilibrium to be reached [21]. The charge efficiency describes the ratio of salt adsorption over charge, for a cycle where the cell voltage $V_{cell}$ is switched periodically between two values, with the high value applied during salt adsorption, and the low value during salt desorption. The low cell voltage is most often $V_{cell}$=0 V, applied by simply electrically short-circuiting the two cells during the salt desorption step. As we will demonstrate in this work, the charge efficiency $\Lambda$ is a very powerful concept to test the suitability of EDL models proposed for ion adsorption in electrified materials.

In a porous carbon material where pores are electrolyte-filled, the surface charge is screened by the adsorption of counterions, and by the desorption of co-ions. The ratio between counterion adsorption, and co-ion desorption, depends on the surface charge. At low surface charge the ratio is one, since for each pair of electrons transferred to a carbon particle, one cation is adsorbed and one anion is expelled, a phenomenon called "ion swapping" by Wu *et al.* [22]. This local conservation of the total ion density is a general feature of the linear response of an electrolyte to an applied voltage smaller than the thermal voltage [23-25]. When in the opposite electrode the same occurs, the charge efficiency $\Lambda$ of the electrode pair will be zero: there is no net salt adsorption from the electrolyte solution flowing in between the two electrodes. In the opposite extreme of a very high surface charge we approach the limit where counterion adsorption is responsible for 100% of the charge screening and we come closer to the limit of $\Lambda$=1 where for each electron transferred between the electrodes one salt molecule is removed from the solution flowing in between the electrodes [7]. This regime exemplifies the strongly nonlinear response of an electrolyte to a large voltage, greatly exceeding the thermal voltage [23, 24]. Correct prediction of the charge efficiency is one requirement of a suitable EDL model.

Contrary to what has often been reported over the past decades [2, 26-32], it is not the mesopores (2-50 nm), but the micropores (< 2 nm), that are the most effective in achieving a high desalination capacity by CDI [33, 34]. Interestingly, already in 1999, based on capacitance measurements in 30% $H_2SO_4$ solutions, Lin *et al.* [35] identified the pore range 0.8-2 nm as the optimum size for EDL formation. The microporous activated carbon MSP-20 (micropore volume 0.96 mL/g, 98 % of all pores are microporous), has the highest reported desalination capacity, of 16.8 mg/g (per g of active material), when tested at 5 mM NaCl and a 1.2 V cell voltage [34, 36]. In these micropores, typically the Debye length $\lambda_D$ will be of the order of, or larger than, the pore size. In water at room temperature, the length scale $\lambda_D$ (in nm) can be approximated by $\lambda_D \sim 10/\sqrt{c_\infty}$ with $c_\infty$ the salt concentration in mM. This implies that the Debye length is around $\lambda_D \sim 3$ nm for $c_\infty$=10 mM. Note that the Debye length is not based on the salt concentration within the EDL, but on the salt concentration $c_\infty$ outside the region of EDL overlap, thus in the interparticle pores outside the carbon particles. Because of the high ratio of Debye length over pore size, in constructing a simple EDL model, it is a good approach for such microporous materials to assume full overlap of the two diffuse Gouy-Chapman layers [19, 37] extending from each side of the pore, leading to an EDL model based on the Donnan concept, in which the electrical potential makes a distinct jump from a value in the space outside the carbon particles to another value within the carbon micropores, without a further dependence of potential on the exact distance to the carbon walls, see Figure 1b [21, 25, 38-41]. The Donnan approximation is the mathematical limit of the mean-field Poisson-Boltzmann (PB) theory for overlapping diffuse double layers, when the Debye length greatly exceeds pore size. In this limit the exact pore geometry is no longer of importance in PB theory, and neither is the surface area. Instead, only the pore volume matters. As an example of the validity of the Donnan model, for a slit-shaped pore with 1 M volumetric charge density and for an external salt concentration of $c_\infty$=10 mM, for any pore size below 10 nm the charge efficiency $\Lambda$ according to the PB-equation is ~0.98, in exact agreement with the Donnan model. Similar Donnan concepts are used in the field of membrane science [42, 43], polyelectrolyte theory [44, 45] and colloidal sedimentation [46, 47].

To describe experimental data for desalination in CDI, the most simple Donnan approach, where only the jump in potential from outside to inside the pore is considered, must be extended in two ways. First of all, an additional capacitance must be included which is located between the ionic diffuse charge and the electronic charge in the carbon. This capacitance may be due to a voltage drop within the carbon itself (space charge layer, or quantum capacitance) [48]. Another reason can be the fact that the ionic charge and electronic charge cannot come infinitely close, e.g., due to the finite ion size, and a dielectric layer of atomic dimension is located in between, called the Stern layer, see Fig. 1. In this work we describe this additional capacitance using the Stern layer-concept. Secondly, to describe data it was found necessary to include an excess chemical potential, $-\mu_{att}$, that describes an additional attraction of each ion to the micropore. This attraction may result from chemical effects [49, 50], but below we propose a quantitative theory based on electrostatic image forces [51, 52], not captured by the classical mean-field approximation of the Donnan model.

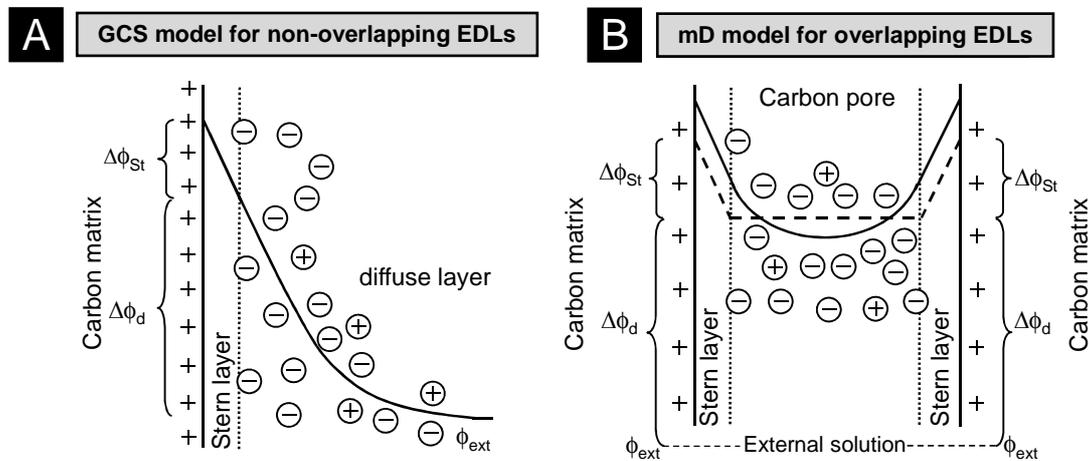

Fig. 1. Schematic view of electrical double layer models used for microporous carbon electrodes. Solid and dashed lines sketch the potential profile, and outside the Stern layers also indicate the profile of counterion concentration. a) Gouy-Chapman-Stern theory for a planar wall without electrical double layer (EDL) overlap. The intersection of Stern layer and diffuse layer is the Stern plane, or outer Helmholtz plane. b) Modified Donnan model. The strong overlap of the diffuse layers (solid line) results in a fairly constant value of diffuse layer potential and ion concentration across the pore (unvarying with pore position), the more so the smaller the pores. In the Donnan model this potential and the ion concentrations are set to a constant value (horizontal dashed line).

Because of its mathematical simplicity, this modified Donnan (mD) model is readily included in a full 1D or 2D porous electrode theory and therefore not only describes the equilibrium EDL structure, but can also be used in a model for the dynamics of CDI [37, 40, 53-55]. Despite this success, it was found that the mD model is problematic when describing simultaneously multiple data sets in a range of values of the external salt concentration. Even when data at just two salt concentrations are simultaneously considered, such as for NaCl solutions with 5 mM and 20 mM salt concentration, it was problematic for the mD model to describe both data sets accurately. Extending the measurement range to 100 mM and beyond, these problems aggravate, see the mismatch between data and theory in Fig. 3 in ref. [38]. We ascribe this problem to the constancy of $\mu_{att}$ in the standard version of the mD model. One extreme consequence of this assumption is that the model predicts unrealistically high salt adsorptions when uncharged carbon is contacted with water of seawater concentration (~0.5 M), e.g. for a typical value of $\mu_{att}$=2.0 kT and $c_\infty$=0.5 M, it predicts an excess salt concentration in uncharged carbons of 3.19 M (with the excess ion concentration twice this value), which is very unrealistic.

In this manuscript we present a physical theory of electrical double layers (EDLs) in microporous carbon electrodes that explains why $\mu_{att}$ is not a constant but in effect decreases at increasing values of micropore total ion concentration. In this way the spurious effect of the prediction of a very high salt adsorption of carbons brought in contact with seawater, is avoided. Section 2 describes the theory together with the implementation of the mD model for CDI. In Section 3 we collect various published data sets of CDI using commercial film electrodes based on activated carbon powders, and fit the data with a simple equation for $\mu_{att}$ inversely dependent on the micropore ion concentration, consistent with our model of image forces. This is a simple theory that has the advantage over more comprehensive and detailed EDL models [22, 56-64] that it can be readily included in a full porous electrode transport theory. Now that $\mu_{att}$ is no longer a constant but a function of the total ion concentration in the pore, which via the Boltzmann relation depends on $\mu_{att}$, a coupled set of algebraic equations results to describe charge and salt adsorption in the micropores of porous carbons. We will demonstrate that across many data sets, this modification improves the predictive power of the mD model very substantially, without predicting extreme salt adsorption at a high salinity anymore.

**2. Theory**

2.1 General description of modified Donnan model

To describe the structure of the EDL in microporous carbons, the modified Donnan (mD) model can be used, which relates the ion concentrations inside carbon particles (in the intraparticle pore space, or micropores, "mi") to the concentration outside the carbon particles (interparticle pore space, or macropores) [65, 66]. At equilibrium, there is no transport across the electrode, and the macropore concentration is equal to that of the external solution outside the porous electrode, which we will describe using the subscript "∞".

In general, in the mD model the micropore ion concentration relates to that outside the pores according to the Boltzmann equilibrium,

$$c_{mi,i} = c_{\infty,i} \cdot \exp\left(-z_i \cdot \Delta\phi_d + \mu_{att,i}\right) \tag{1}$$

where $z_i$ is the valency of the ion, and $\Delta\phi_d$ the Donnan potential, i.e., the potential increase when going from outside to inside the carbon pore. This is a dimensionless number and can be multiplied by the thermal voltage $V_T = RT/F$ to obtain the Donnan voltage with dimension V.

The Donnan voltage is a potential of mean force derived from the Poisson-Boltzmann mean-field theory, which assumes that the electric field felt by individual ions is generated self-consistently by the local mean charge density. Therefore, the excess chemical potential of each ion, $-\mu_{att}$, has a contribution from electrostatic correlations, which is generally attractive if dominated by image forces in a metallic micropore, as described below. We use $\mu_{att}$ as a dimensionless number which can be multiplied by $kT$ to obtain an energy per ion with dimension J.

In the mD model we consider that outside the carbon particles there is charge neutrality,

$$\sum_i z_i \cdot c_{\infty,i} = 0 \tag{2}$$

while inside the carbon micropores, the micropore ionic charge density (per unit micropore volume, dimension mol/m$^3$=mM) is given by

$$\sigma_{mi} = \sum_i z_i \cdot c_{mi,i}. \tag{3}$$

This ionic charge is compensated by the electronic charge in the carbon matrix: $\sigma_{mi} = -\sigma_{elec}$. In using this simple equation we explicitly exclude the possibility of chemical surface charge effects, but such an effect can be included [54]. Eq. 3 describes local electroneutrality in the micropores, a well-known concept frequently used in other fields as well, such as in polyelectrolyte theory [44, 45], ion-exchange membranes [42, 43], and colloidal sedimentation [46, 47]. The ionic charge density relates to the Stern layer potential difference, $\Delta\phi_{St}$, according to

$$\sigma_{mi} = -C_{St,vol} \cdot \Delta\phi_{St} \cdot V_T / F \tag{4}$$

where $C_{St,vol}$ is a volumetric Stern layer capacity in F/m$^3$. For $C_{St,vol}$ we use the expression

$$C_{St,vol} = C_{St,vol,0} + \alpha \cdot \sigma_{mi}^2 \tag{5}$$

where the second term empirically describes the experimental observation that the Stern layer capacity goes up with micropore charge, where $\alpha$ is a factor determined by fitting the model to the data [33, 67-69]. To consider a full cell we must add to Eqs. 1-5 (evaluated for both electrodes) the fact that the applied cell voltage relates to the EDL voltages in each electrode according to

$$V_{cell}/V_T = \left|\Delta\phi_d + \Delta\phi_{St}\right|_{cathode} + \left|\Delta\phi_d + \Delta\phi_{St}\right|_{anode}. \tag{6}$$

Allowing for unequal electrode mass, we have as an additional constraint that the electronic charge in one electrode plus that in the other, sum up to zero,

$$\sigma_{mi,cathode} \cdot mCmA = -\sigma_{elec,cathode} \cdot mCmA = \sigma_{elec,anode} = -\sigma_{mi,anode} \tag{7}$$

where mCmA is the mass ratio cathode-to-anode. In an adsorption/desorption cycle, the adsorption of a certain ion i by the cell pair, per gram of both electrodes combined, is given by [39]

$$\Gamma_i = \upsilon_{mi} \cdot \left[\frac{mCmA}{mCmA+1} \cdot \left(c_{mi,i}^{cathode} - c_{mi,i}^0\right) + \frac{1}{mCmA+1} \cdot \left(c_{mi,i}^{anode} - c_{mi,i}^0\right)\right] \tag{8}$$

where superscript "0" refers to the discharge step, when typically a zero cell voltage is applied between anode and cathode. In Eq. 8, $\upsilon_{mi}$ is the micropore volume per gram of electrode which in an electrode film is the product of the mass fraction of porous carbon in an electrode, e.g. 0.85, and the pore volume per gram of carbon, as measured for instance by $N_2$ adsorption analysis. The question of which pore volume to use (i.e., based on which pore size range) is an intricate question addressed in ref. [34].

This set of equations describes the mD model for general mixtures of ions, and includes the possibility of unequal electrode masses (e.g., a larger anode than cathode) and unequal values for $\mu_{att}$ for the different ions. For the specific case of a 1:1 salt as NaCl, the cation adsorption equals the anion adsorption, and thus Eq. 8 also describes the salt adsorption, $\Gamma_{salt}$, in a cycle. The charge per gram of both electrodes $\Sigma$ is given by $\Sigma = \upsilon_{mi}/(mCmA+1) \cdot \left(\sigma_{mi}^{anode} - \sigma_{mi}^{anode,0}\right)$. The ratio of these two numbers is the charge efficiency of a CDI cycle, $\Lambda$, see Fig. 5, for various values of mCmA.

### 2.3 Equal electrode mass

Next we limit ourselves to the case that there is an equal mass of anode and cathode, i.e., the two electrodes are the same, and thus $\sigma_{elec,cathode}+\sigma_{elec,anode}=0$. After solving Eqs. 1-5 for each electrode separately, together with Eq. 6, we can calculate the electrode charge, and salt adsorption by the cell.

Multiplying micropore charge density $\sigma_{mi}$ (for which we can take any of the values considered in Eq. 7, ionic or electronic, in the anode or in the cathode, as they are all the same when mCmA=1) by Faraday's constant, $F$, and by the volume of micropores per gram of electrode, $\upsilon_{mi}$, we obtain for the charge $\Sigma_F$ in C/g,

$$\Sigma_F = \tfrac{1}{2} \cdot F \cdot \upsilon_{mi} \cdot \left|\sigma_{mi} - \sigma_{mi}^0\right| \tag{9}$$

and for the ion adsorption of a cell pair,

$$\Gamma_i = \tfrac{1}{2} \cdot \upsilon_{mi} \cdot \left(c_{mi,i}^{cathode} - c_{mi,i}^{cathode,0} + c_{mi,i}^{anode} - c_{mi,i}^{anode,0}\right). \tag{10}$$

In case the cell voltage is set to zero during discharge, then (without chemical charge on the carbon walls) $\sigma_{mi,i}^0 = 0$ and $c_{mi,i}^{cathode,0} = c_{mi,i}^{anode,0} = c_{mi,i}^0$.

### 2.4 Monovalent salt solution – equal electrode mass

Next we focus on a 1:1 salt such as NaCl, in addition to assuming that the two electrodes have the same mass. In the case of a 1:1 salt, $c_{\infty,cation}$ is equal to $c_{\infty,anion}$ and we can equate both to the external salt concentration, $c_\infty$. From this point onward, we will assume $\mu_{att}$ to be the same for $Na^+$ as $Cl^-$ [see note 1 at end of manuscript]. For a 1:1 single-salt solution, combination of Eqs. 1-5 leads to

$$\sigma_{mi} = c_{cation,mi} - c_{anion,mi} = -2 \cdot c_\infty \cdot \exp(\mu_{att}) \cdot \sinh(\Delta\phi_d) \tag{11}$$

and

$$c_{ions,mi} = c_{cation,mi} + c_{anion,mi} = 2 \cdot c_\infty \cdot \exp(\mu_{att}) \cdot \cosh(\Delta\phi_d). \tag{12}$$

Because of symmetry, in this situation Eq. 6 simplifies to

$$V_{cell}/V_T = 2 \cdot \left|\Delta\phi_d + \Delta\phi_{St}\right| \tag{13}$$

while only one electrode needs to be considered. For a 1:1 salt, the amount of anion adsorption by the cell pair equals the amount of cation adsorption, and thus $\Gamma_{cation}=\Gamma_{anion}=\Gamma_{salt}$.

The charge efficiency, being the measurable equilbrium ratio of salt adsorption $\Gamma_{salt}$ over charge $\Sigma$ is now given by (clearly defined as an integral quantity)

$$\Lambda = \frac{\Gamma_{salt}}{\Sigma} = \frac{c_{ions,mi} - c^0_{ions,mi}}{|\sigma_{mi}|} = \tanh\frac{|\Delta\phi_d|}{2} \quad (14)$$

in case that 1. the reference condition (condition during ion desorption-step) is a zero cell voltage, and 2. we use the single-pass method of testing, where the salt concentration $c_\infty$ is the same before and after applying the voltage. Note that this condition does not apply when the batch mode of CDI testing is used where $c_\infty$ is different between the end of the charging and the end of the discharging step (see ref. [7]). In Eq. 14 the charge $\Sigma$, expressed in mol/g, is equal to $\Sigma_F$ divided by $F$. Eq. 14 demonstrates that $\Lambda$ is not directly dependent of such parameters as $\mu_{att}$, $C_{St,vol}$ or $c_\infty$, but solely depends on $\Delta\phi_d$ [16, 23]. Of course, in an experiment with a certain applied cell voltage, all of these parameters do play a role in determining the value of $\Lambda$ via their influence on $\Delta\phi_d$. An equation similar to Eq. 14 is given in the context of ion transport through lipid bilayers as Eqs. 8 and 10 in ref. [70].

2.5 Simple Theory of Image Forces in Micropores

Here, we propose a first approximation of $\mu_{att}$ due to image forces between individual ions in the micropores and the metallic carbon matrix [71], leading to a simple formula that provides an excellent description of our experimental data below. Image forces have been described with discrete dipole models for counterion-image monolayers [52], as well as (relatively complicated) extensions of Poisson-Boltzmann theory [51]. Simple modified Poisson equations that account for ion-ion Coulomb correlations that lead to charge oscillations in single component plasmas [72, 73] and multicomponent electrolytes or ionic liquids [74] are beginning to be developed, but image forces at metallic or dielectric surfaces have not yet been included. Moreover, to our knowledge, image forces have never been included in any mathematical model for the dynamics of an electrochemical system.

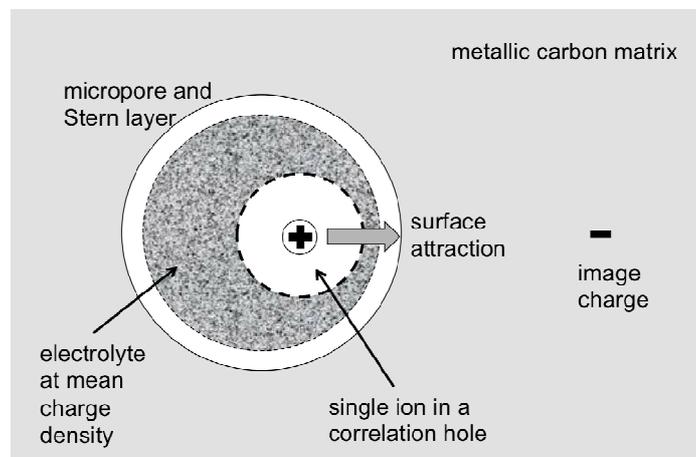

Fig. 2. Sketch of electrostatic image correlations leading to attractive surface forces for all ions in a micropore, whose size is comparable to both the Bjerrum length and the mean ion spacing.

Consider a micropore of size $\lambda_p \approx$ 1-5 nm whose effective permittivity, $\varepsilon_p$, is smaller than that of the bulk electrolyte, $\varepsilon_b$, due to water confinement and the dielectric decrements of solvated ions. The local Bjerrum length in the micropore,

$$\lambda_B = \frac{e^2}{4\pi\varepsilon_p kT} \tag{15}$$

is larger than its bulk value ($\lambda_B$=0.7 nm in water at room temperature) by a factor of $\varepsilon_b/\varepsilon_p \approx$ 5-10 and thus is comparable to the pore size. As such, ions have strong attractive Coulomb interactions $-E_{im} > kT$ with their image charges from anywhere within the micropore. For a spherical metallic micropore of radius $\lambda_p$, the image of an ion of charge $q=\pm ze$ at radial position $r$ has charge $\bar{q} = \mp \lambda_p q / r$ and radial position $\bar{r} = \lambda_p^2 / r$ outside the pore [75], see Fig. 2. The attractive Coulomb energy between the ion and its image,

$$E_{im}(r) = \frac{\lambda_p (ze)^2}{4\pi\varepsilon_p (\lambda_p^2 - r^2)} \tag{16}$$

diverges at the surface, $r \rightarrow \lambda_p$, but the Stern layer of solvation keeps the ions far enough away to prevent specific adsorption. For consistency with the Donnan model, which assumes constant electrochemical potentials within the micropores (outside the Stern layers), we approximate the image attraction energy by a constant, equal to its value at the center of the micropore,

$$E_{im} \approx \frac{(ze)^2}{4\pi\varepsilon_p \lambda_p} = z^2 \cdot kT \cdot \frac{\lambda_B}{\lambda_p} \ . \tag{17}$$

This scaling is general and also holds for other geometries, such as parallel-plate or cylindrical pores, with a suitable re-definition of $\lambda_p$. The image force on a given ion is significantly reduced by the presence of other ions due to Coulomb correlations, which effectively converts the bare ion monopole into collections of fluctuating multipoles with more quickly decaying electric fields. The attractive excess chemical potential, $\mu_{att} = E_{im} P_{im}$, is thus multiplied by the probability that an ion falls into a "correlation hole," or fluctuating empty region, and feels a bare image force. If the mean volume of a correlation hole, $c_{ions,mi}^{-1}$, is smaller than the characteristic pore volume, $\lambda_p^3$, then the probability that a given particle enters a correlation hole scales as $P_{im} \approx (\lambda_p^3 c_{ions,mi})^{-1}$. This implies that the excess chemical potential due to image forces is inversely proportional to the *total* concentration of all ions, since the image energy is independent of the sign of the charge.

We thus arrive at a very simple formula for the excess attractive chemical potential

$$\mu_{att} = \frac{E}{c_{ions,mi}} \tag{18}$$

where

$$E = z^2 \cdot kT \cdot \lambda_B \cdot \lambda_p^{-4} \ . \tag{19}$$

## 3. Results and Discussion

In sections 3.1-3.3 we present a re-analysis of three sets of data of water desalination by CDI using commercial composite carbon film electrodes. These data were previously compared with the standard mD model (that assumes $\mu_{att}$ to be a constant). Here we will demonstrate how making $\mu_{att}$ a simple function of $c_{ions,mi}$ according to Eq. (18) significantly improves the fit to the data, without extra fitting parameters. This we call the improved mD model. In the last section 3.4 we analyse experiments of salt adsorption in uncharged carbon to have direct access to the energy parameters $E$ and $\mu_{att}$ and also include ion and pore wall volume effects by using a modified Carnahan-Starling equation of state.

In all sections 3.1-3.3 the same parameter settings are used, being $p_{mi}$=0.30 and $\rho_{elec}$=0.55 g/mL, thus $\upsilon_{mi}=p_{mi}/\rho_{elec}$=0.545 mL/g, $C_{St,vol,0}$=0.145 GF/m$^3$, $\alpha$=30 F·m$^3$/mol$^2$, $E$=300 kT·mol/m$^3$. The carbon electrodes used in sections 3.1-3.3 are based on a commercial material provided by Voltea B.V. (Sassenheim, The Netherlands) which contained activated carbon, polymer binder and carbon black. This material was used in all our studies in refs. [21, 25, 38, 39, 76, 77].

### 3.1 Data for varying cell voltage at two values of salt concentration (5 and 20 mM NaCl)

Data for charge and salt adsorption by a symmetric pair of activated carbon electrodes as function of salt concentration (5 and 20 mM NaCl) and cell voltage was presented in ref. [16] and was re-analyzed using the standard mD model in ref. [38]. By "standard" we imply using a fixed value of $\mu_{att}$. Though a reasonable good fit was obtained, see Fig. 2b in ref. [38], the effect of salinity $c_\infty$ was overestimated.

To describe in more detail how well the standard mD model fits the data, we make the following analysis: As Eq. 11 demonstrates, according to the standard mD approach, there is a direct relationship between the ratio $\sigma_{mi}/c_\infty$ and $\Delta\phi_d$, and thus, according to Eq. 14, there is also a direct relationship between $\sigma_{mi}/c_\infty$ and $\Lambda$. Thus, two datasets (each for a range of cell voltages) obtained at two values of the external salinity, $c_\infty$, should overlap. However, as Fig. 3a demonstrates, the two datasets do not, and stay well separated. This is direct evidence that the standard mD model with a fixed $\mu_{att}$ is not accurate enough. A direct check whether a modified mD model works better, is to plot the two datasets together with the corresponding two modeling lines (thus for two values of $c_\infty$), all in one graph, and choose such an x-axis parameter that the modeling lines overlap, and check if now the two datasets overlap better. This procedure is followed in Fig. 3b where it is clearly observed that when we plot $\Lambda$ vs. $\sigma_{mi}/(c_\infty/c_{ref})^a$ with $c_{ref}$=20 mM and the power $a$ equal to $a$=0.31, the modeling lines for the improved mD model collapse on top of one another, and also the data now almost perfectly overlap. Note that the value of $a$=0.31 has no special significance as far as we know, it is just a chosen value to make the modeling lines overlap. Clearly, the use of the improved mD model to describe $\mu_{att}$ as function of $c_{ions,mi}$ results in a significantly better fit of the model to the data.

Fig. 3c plots the total ion concentration in the micropore volume vs. the micropore charge. In this representation we observe again a good fit of the improved mD model to the data. Note that the modeling fit in Fig. 3b and 3c is independent of details of the Stern layer (see Eq. 5), and only depends on the value of $E$ and the micropore volume, $v_{mi}$, see Eqs. 11 and 12. The deviation from the 100% *integral* charge efficiency $\Lambda$-line (dashed line at angle of 45 degrees) is larger for 20 mM than for 5 mM.

Note that for both salt concentrations there is a range where the data run parallel to this line. In this range, beyond a micropore charge density of ~200 mM, the *differential* charge efficiency $\lambda$ is unity and if we would stay in this (voltage) range, for each electron transferred between the electrodes, we remove one full salt molecule. The parallelism of these two lines is typical of EQCM response of carbons, observed for moderate charge densities [78]. Fig. 3d is the classical representation of $\Lambda$ vs $V_{cell}$, and we obtain a much better fit than in Fig. 2b in ref. [38], with the influence of the external salinity no longer overpredicted. Furthermore, we reproduce in Fig 3e-3g the direct measurement data of salt adsorption in mg/g and charge in C/g for this material, and find a very good fit of the model to these four data sets [see note 2]. In Fig. 3h we recalculate data and theory to the counterion and coion concentrations in the pores, a graph similar to one by Oren and Soffer [13] and by Kastening *et al.* [65, 79]. Analysing the model results, e.g. for $c_\infty$=5 mM, the predicted total ion concentration in the pores increases from a minimum value of $c_{ions,mi}$=120 mM at zero charge, to about 1000 mM at $V_{cell}$=1.4 V.

With a value of $E$=300 kT·mol/m$^3$, at $c_\infty$=5 mM the attraction energy, $\mu_{att}$, is at a maximum of 2.48 kT at zero charge, and decreases steadily with charging, to a value of $\mu_{att}$=0.28 kT at $V_{cell}$=1.4 V. The attractive energy inferred from the experimental data using the mD porous electrode model is quantitatively consistent with Eq. (19) from our simple theory of image forces without any fitting parameters. Using $\lambda_p = 2$ nm, the experimental value $E$=0.3 $kT$·M implies $\lambda_B$=2.9 nm, or $\varepsilon_p = 0.25\varepsilon_b = 20\varepsilon_0$, which is a realistic value for the micropore permittivity. Admittedly, the quantitative agreement may be fortuitous and could mask other effects, such as ion adsorption equilibria, but it is clear that the overall scale and concentration dependence of the attractive energy are consistent with image forces.

Fig. 3. Analysis of data of salt adsorption and charge efficiency $\Lambda$ for NaCl solutions at $c_\infty$=5 and 20 mM for cell voltages up to $V_{cell}$=1.4 V. a) Plotting $\Lambda$ vs. the ratio $\sigma_{mi}/c_\infty$ does not lead to overlap of the datasets, demonstrating that using $\mu_{att}$=constant in the standard mD model is not correct. b) Plotting $\Lambda$ vs $\sigma_{mi}\cdot(c_{ref}/c_\infty)^a$ with $a$=0.31 leads to a perfect overlap of the modeling lines, where $\mu_{att}$=$E/c_{ions,mi}$ and $E$=300 kT·mol/m$^3$. The datasets now also overlap quite closely demonstrating the relevance of the use of the improved mD model ($c_{ref}$=20 mM). c) Using the improved mD model, the total excess micropore ion adsorption, $c_{ions,mi}$ (equal to salt adsorption in a symmetric cell pair), is plotted vs micropore charge density, $\sigma_{mi}$, showing the expected deviation from the 100% charge efficiency-line. d) Theory of $\Lambda$ vs $V_{cell}$ according to the improved mD model (compared with data), showing a much smaller influence of $c_\infty$ on $\Lambda$ than in the standard mD model, see Fig. 2b in ref. [38]. e)-g) Direct data of salt adsorption $\Gamma_{salt}$ in mg/g and charge $\Sigma_F$ in C/g, compared with the improved mD model. g) Calculated micropore ion concentrations as function of electrode charge, again compared with the improved mD model.

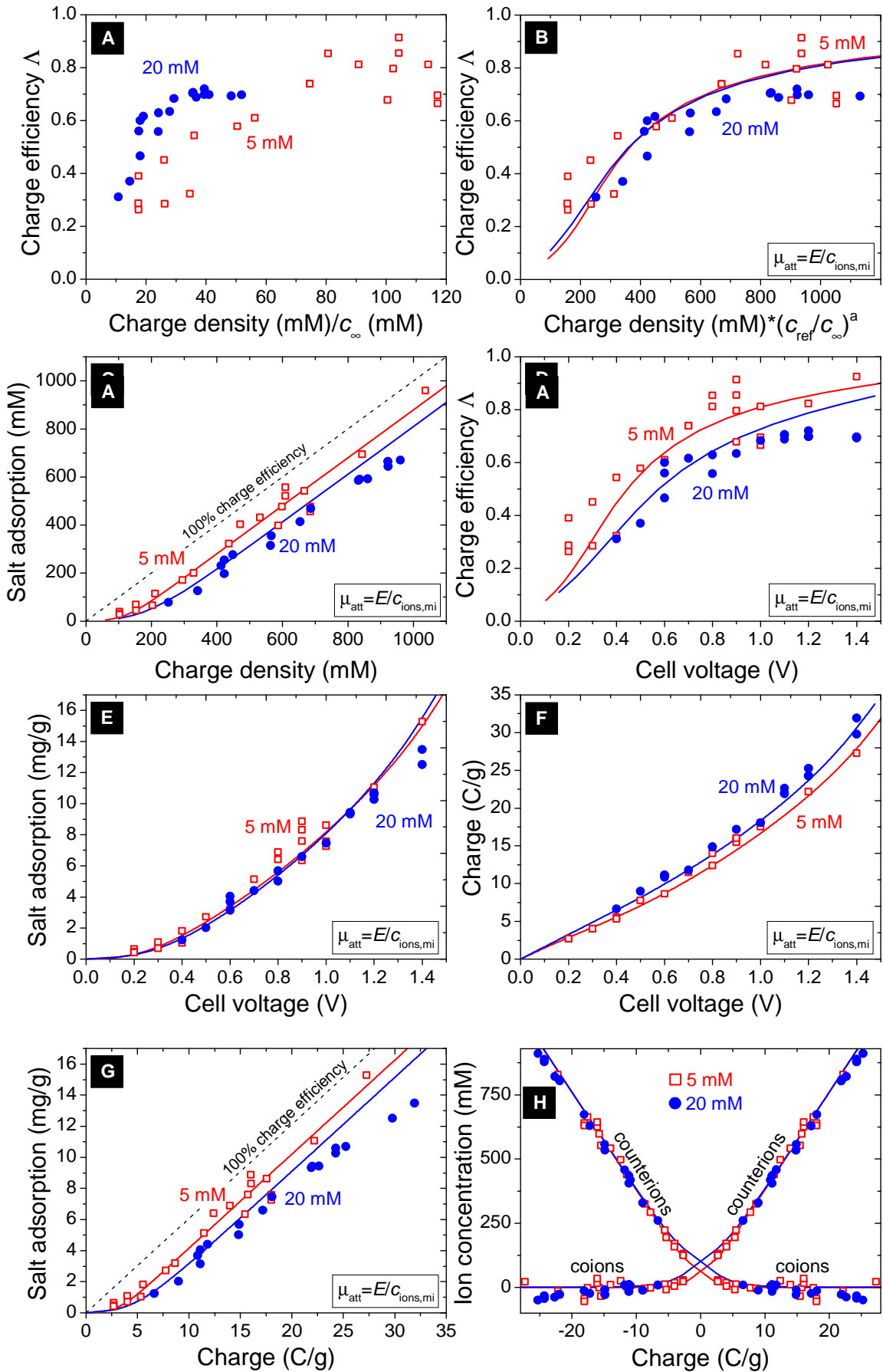

## 3.2 Data at one cell voltage level for a range of salt concentrations (2.5-200 mM NaCl)

Next, we extend the testing of the same improved mD model to a much larger range of NaCl salt concentrations, from 2.5 to 200 mM, all evaluated at 1.2 V cell voltage, see Fig 3 of ref. [38] for NaCl salt concentrations ranging from 2.5 mM to 100 mM. In Fig. 4 this range is extended to 200 mM by including extra unpublished work related to material in refs. [21, 77]. In ref. [38] using the standard mD model it proved impossible to fit the model to the data for the whole range of salinities, with beyond $c_\infty$=40 mM the charge underestimated, and salt adsorption underestimated even more, leading to an underprediction of the charge efficiency, see the line denoted "$\mu_{att}$=constant" in Fig. 4b. The experimental observation that the salt adsorption does not change much with external salt concentration up to 100 mM, could not be reproduced at all. However, with the modification to make $\mu_{att}$ inversely proportional to $c_{ions,mi}$, a very good fit to the data is now obtained, both for charge and for salt adsorption, as we can observe in Fig. 4a. Fig. 4b presents results of the charge efficiency $\Lambda$, which is the ratio of salt adsorption to charge (including Faraday's number to convert charge to dimension mol/g), and as can be observed, the improved mD model using $\mu_{att}=E/c_{ions,mi}$ shows a much better fit to the data than the standard mD model which assumes $\mu_{att}$ to be constant.

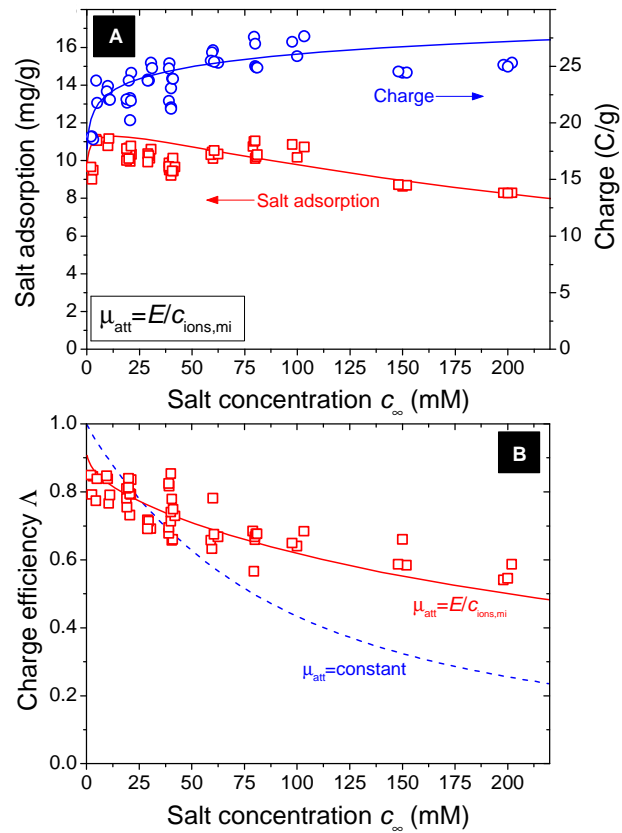

Fig. 4. a) Salt adsorption and charge density, and b) charge efficiency $\Lambda$, for CDI in a range of NaCl salt concentrations (2.5-200 mM) at $V_{cell}$=1.2 V. Both in a) and b) solid lines denote calculation results of the improved mD model while in b) the dashed line is based on the standard mD model.

3.3 Data for unequal electrode mass (5 and 20 mM NaCl)

Data for NaCl adsorption in asymmetric CDI systems were presented by Porada *et al.* [39] That work is based on varying the electrode mass ratio, i.e., by placing on one side of the spacer channel two or three electrodes on top of one another. In this way a cell is constructed which has two times, or three times, the anode mass relative to cathode mass, or vice versa. In Fig. 5 we present the data for charge efficiency, $\Lambda$, defined as salt adsorption by the cell pair divided by charge, vs the mCmA ratio, which is the mass ratio of cathode to anode. Here data are presented at a cell voltage of $V_{cell}$=1.0 V and a salt concentration of 5 and 20 mM, like Fig. 3c in ref. [39]. Comparing with the fit obtained by Porada *et al.* using a constant value of $\mu_{att}$ (dashed lines in Fig. 5), a significantly improved fit is now achieved.

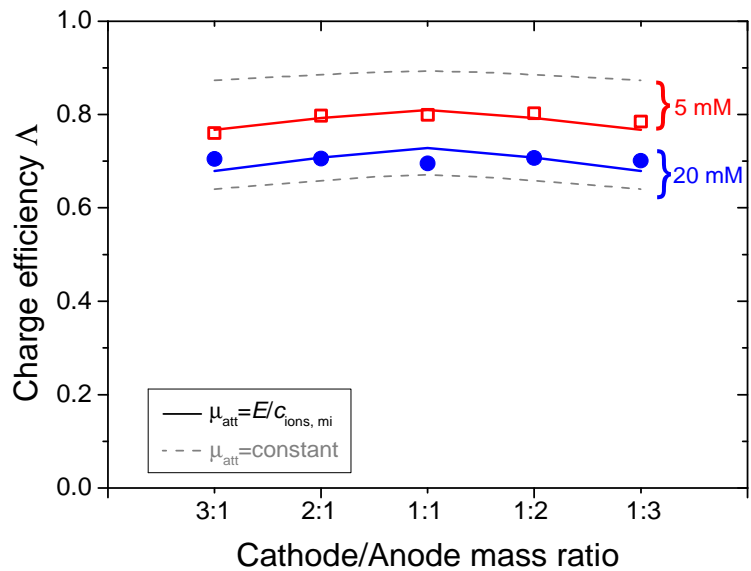

Fig. 5. Charge efficiency for CDI at unequal mass ratio cathode-to-anode ($V_{cell}$=1.0 V). Dashed lines show prediction using a fixed $\mu_{att}$ in the standard mD model, while the solid lines show results for the improved mD model where $\mu_{att}=E/c_{ions,mi}$.

3.4 Analysis of data for adsorption of salt in uncharged carbon – measuring $\mu_{att}$

A crucial assumption in the mD models is the existence of an attractive energy, $\mu_{att}$, for ions to move into carbon micropores, which we attribute to image forces as a first approximation. The existence of this energy term implies that uncharged carbons must adsorb some salt, as known from refs. [49, 50] and references therein, and as can also be inferred from refs. [65, 80]. In the present section we show results of the measurement of $\mu_{att}$ by directly measuring the adsorption of NaCl in an activated carbon powder (Kuraray YP50-F, Kuraray Chemical, Osaka, Japan). This carbon is mainly microporous with 0.64 mL/g in the pore size range <2 nm and 0.1 mL/g mesopores [34].

The carbon powder was washed various times in distilled water and filtered, to remove any possible ionic/metallic constituents of the carbon, and was finally dried in an oven at 100 °C. A volume of water $V$ with pre-defined NaCl concentration was mixed with various amounts of carbon (mass $m$) in sealed flasks. These flasks were gently shaken for 48 hours. The carbon/water slurry is pressed through a Millipore Millex-LCR filter (Millipore, Massachusetts, USA) and the supernatant was analyzed to measure the decrease in salt concentration, from which we calculate the salt adsorption. Note that both the initial and final salt concentration in the water are analyzed in the same analysis program, and the difference is used to calculate salt adsorption. The pH that initially was around pH 6-7 increased to values pH~9 after soaking with carbon. By IC (ion chromatography) we measure the Cl$^-$-content of the supernatant [see note 3].

The excess salt adsorption is calculated from the measured decrease in Cl$^-$-concentration, $\Delta c$, in in the supernatant (relative to that in the initial solution) according to $n_{salt,exc}=V/m \cdot \Delta c$ in mol/g, which we multiply by $\upsilon_{mi}$=0.64 mL/g to obtain an estimate for the excess salt concentration in the pores, $c_{exc}$. The excess concentration is plotted against the final (after equilibration) salt concentration (again based on the measured Cl$^-$-concentration) in Fig. 6. By "excess concentration" we mean the difference in concentration in the carbon pores, relative to that outside the carbon particles.

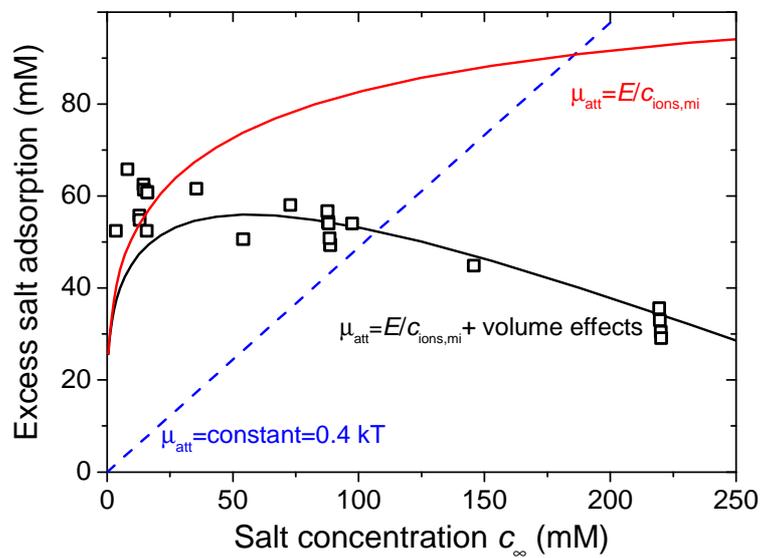

Fig. 6. Excess adsorption of NaCl in uncharged carbon as function of external salt concentration, $c_\infty$. Straight dashed line based on standard mD model (constant $\mu_{att}$=0.4 kT), upper curved line based on improved mD model as used in sections 3.1-3.3, and lower curved line for extended model including ion-volume effects.

As Fig. 6 demonstrates, the measured excess salt adsorption, $c_{exc}$, as function of the external salt concentration, $c_\infty$, has a broad maximum in the range from 5 to 100 mM, and beyond that $c_{exc}$ decreases gradually. The measured excess adsorption of around 55 mM recalculates to a salt adsorption of about 2 mg/g. This number is about a factor of 10 lower than values obtained for mixed adsorbents (some containing activated carbon) reported in ref. [81] and a factor 5 lower than values for alkali and acid adsorption reported by Garten and Weiss [50]. The standard mD model assuming a constant $\mu_{att}$ does not describe these data at all. Here in Fig. 6 is plotted a line for $\mu_{att}$=0.4 kT, much lower than values for $\mu_{att}$ used by us in earlier work, which were mostly around $\mu_{att}$=1.5 kT, but even this low value of $\mu_{att}$=0.4 kT results in a model prediction which significantly overpredicts $c_{exc}$ at salt concentrations beyond $c_\infty$=100 mM. Thus, taking a constant value of $\mu_{att}$ does not describe data at all. The improved mD model using $\mu_{att}=E/c_{ions,mi}$ (upper curved line) works much better and more closely describes the fact that $c_{exc}$ levels off with increasing $c_\infty$. However, it still does not describe the fact that a maximum develops, and that at high $c_\infty$ the excess adsorption decreases again. To account for this non-monotonic effect we include an ion volume-correction according to a modified Carnahan-Starling equation-of-state [44, 45, 47, 63]. For an ion, because of its volume there is an excess, volumetric, contribution to the ion chemical potential, both in the external solution and in the micropores. This excess contribution is calculated from

$$\mu_{exc} = \phi \cdot \left(8 - 9 \cdot \phi + 3 \cdot \phi^2\right) \cdot (1-\phi)^{-3} = (3-\phi) \cdot (1-\phi)^{-3} - 3 \tag{20}$$

where $\phi$ is the volume fraction of all ions together. In external solution, to calculate $\mu_{exc}^\infty$, we use $\phi = 2 \cdot v_{ion} \cdot c_\infty$, where $v$ is the ion volume ($v_{ion} = \pi/6 \cdot d_{ion}^3$, where $d_{ion}$ is the ion size) and the factor 2 stems from the fact that a salt molecule consists of two ions, while in the carbon pores, to calculate $\mu_{exc}^{pore}$, we replace in Eq. (20) $\phi$ by $\phi_{eff}$, for which we use the empirical expression $\phi_{eff} = v \cdot c_{ions,mi} + \alpha \cdot d_{ion}/d_{pore}$, derived from fitting calculation results of the average density of spherical particles in a planar slit based on a weighted-density approximation [82, 83], where $\alpha$ is an empirical correction factor of $\alpha = 0.145$ and where $d_{pore}$ is the pore width. For a very large pore, or for very small ions, the correction factor tends to zero.

The excess salt concentration as plotted in Fig. 6 is given by $c_{exc} = \frac{1}{2} \cdot c_{ions,mi} - c_\infty$, and is calculated via

$$c_{ions,mi} = 2 \cdot c_\infty \cdot \exp\left(\mu_{att} + \mu_{exc}^\infty - \mu_{exc}^{pore}\right). \tag{21}$$

Eqs. (20)+(21) presents a self-consistent set of equations that can be solved to generate the curves in Fig. 6. To calculate the lines for the improved mD model without ion volume effects, $\phi$ is set to zero.

To obtain the best fit in Fig. 6, we use $E=220$ kT·mol/m$^3$ and for the ion and pore sizes we use $d_{ion}=0.5$ nm and $d_{pore}=2.5$ nm. This pore size is representative for the microporous material used, while the $E$-value is close to that used in sections 3.1-3.3. For the model including volume effects, the derived values for the term $\mu_{att}$ decrease from $\mu_{att}=2.47$ kT at $c_\infty=5$ mM NaCl to $\mu_{att}=0.46$ kT at 200 mM. The volume exclusion term, $\mu_{exc}^{pore} - \mu_{exc}^\infty$, is quite independent of $c_\infty$ at around 0.28 kT. As can be observed in Fig. 6, beyond 25 mM a good fit is now obtained.

Thus, we conclude that the analysis presented in this section underpins the fact that uncharged carbon absorbs salt, and that the data are well described by a model using an attractive energy term $\mu_{att}$ inversely proportional to the total ion concentration to account for image forces, in combination with a correction to include ion volume effects as an extra repelling force which counteracts ion adsorption at high salt concentrations.

## 4. Conclusions

We have demonstrated that to describe charge and salt adsorption in porous carbon electrodes for capacitive deionization (CDI), that the predictive power of the modified Donnan model can be significantly increased by assuming that the ion attractive energy $\mu_{att}$ is no longer a fixed constant, but is inversely related to the total ion concentration in the pores. In this way, the anomaly of predicted extremely high salt adsorptions in carbons in contact with high-salt solutions such as seawater, is resolved. Whereas in the standard mD model, using as an example a constant value of $\mu_{att}$=2.0 kT, the excess adsorption of salt from water of a salinity of 0.5 M (sea water) into uncharged carbon is predicted to be 3.19 M, in the improved mD model, with $E$=300 kT·mol/m$^3$, this excess adsorption is only 0.13 M, a much more realistic value. Actually, extrapolation of data presented in Fig. 6 suggests that in a 0.5 M salt solution, the effect of ion volume exclusion may be high enough that instead of an excess adsorption, we have less salt in the pores than in the outside solution. The improved Donnan model not only has relevance for modeling the EDL structure in porous electrodes for CDI, but also for membrane-CDI [84-87], salinity gradient energy [40, 55] and for energy harvesting from treating $CO_2$ containing power plant flue gas [88, 89].

This work also highlights for the first time the important role of electrostatic image forces in porous electrodes, which cannot be described by classical mean-field theories. We propose a simple approximation that works very well for the data presented here. This result invites further systematic testing and more detailed theory. The theory can be extended for larger pores with non-uniform ion densities, and by making use of more accurate models of the Stern layer, including its dielectric response and specific adsorption of ions. Unlike the situation for biological molecules [90] and ion channels [91], where image forces are repulsive due to the low dielectric constant, metallic porous electrodes generally exert attractive image forces on ions that contribute to salt adsorption, even at zero applied voltage.

**Notes**

1. Note that for a symmetric 1:1 salt, and for a symmetric electrode, there is no effect of explicitly considering the two values of $\mu_{att,j}$ to be different, as long as their average is same. When considering $\mu_{att,Na}$ and $\mu_{att,Cl}$ to be different, still the same model output (charge and salt adsorption versus cell voltage) is generated and only the individual Donnan potentials change in both electrodes, one up, one down, with their sum remaining the same. However, for asymmetric electrodes, or asymmetric salts (such as $CaCl_2$), and for salt mixtures [25] there is an effect of the individual values of $\mu_{att,i}$ on the measurable performance of a CDI cell. Of course, it must be the case that $\mu_{att}$ differs between different ion types as it is known that specific adsorption of ions increases with their size, which is correlated with their lower solvation ability: from F$^-$ to I$^-$, and from Li$^+$ to Cs$^+$ [92, 93].
2. Note that to analyze the data of ref. [16], as presented in Fig. 2, the mass as assumed erroneously in ref. [16] to be 10.6 g must be corrected to a mass of 8.5 g, and thus the reported salt adsorption and charge in ref. [16] is multiplied by 10.6/8.5; Note that in ref. [38] a correction to 8.0 g was assumed in Fig. 2 there.
3. Also the Na$^+$-concentration was measured in all samples, using Inductively Coupled Plasma mass spectrometry. Analysis of the electrolyte solution prior to contacting with carbon gave a perfect match of Na$^+$-concentration to Cl$^-$-concentration. However, analysis of the supernatant that had been in contact with the carbon quite consistently gave a lower Cl$^-$-concentration than Na$^+$-concentration, by 1-6 mM (thus more Cl$^-$-adsorption in the carbon), in line with the higher reported anion vs cation adsorption for carbons activated beyond 600 $^o$C [50], and for mixed adsorbent samples reported in Fig. 2 of ref. [81].

**Acknowledgments**

Part of this work was performed in the cooperation framework of Wetsus, centre of excellence for sustainable water technology (www.wetsus.nl). Wetsus is co-funded by the Dutch Ministry of Economic Affairs and Ministry of Infrastructure and Environment, the European Union Regional Development Fund, the Province of Fryslân, and the Northern Netherlands Provinces. The authors like to thank the participants of the research theme Capacitive Deionization for fruitful discussions and financial support. We thank Michiel van Soestbergen for providing unpublished theoretical results used in section 3.4.

**References**

1. Arnold, B.B. and G.W. Murphy (1961) Journal of Physical Chemistry 65: 135-138
2. Farmer, J.C., D.V. Fix, G.V. Mack, R.W. Pekala, and J.F. Poco (1996) Journal of the Electrochemical Society 143: 159-169
3. Johnson, A.M. and J. Newman (1971) Journal of The Electrochemical Society 118: 510-517
4. Soffer, A. and M. Folman (1972) Journal of Electroanalytical Chemistry and Interfacial Electrochemistry 38: 25-43
5. Suss, M.E., T.F. Baumann, W.L. Bourcier, C.M. Spadaccini, K.A. Rose, J.G. Santiago, and M. Stadermann (2012) Energy & Environmental Science 5: 9511-9519
6. Rica, R.A., R. Ziano, D. Salerno, F. Mantegazza, and D. Brogioli (2012) Physical Review Letters 109: 156103
7. Porada, S., R. Zhao, A. van der Wal, V. Presser, and P.M. Biesheuvel (2013) Progress in Material Science 58 1388-1442
8. Jeon, S.-I., H.-R. Park, J.-G. Yeo, S. Yang, C.H. Cho, M.H. Han, and D.-K. Kim (2013) Energy & Environmental Science 6: 1471–1475
9. Jande, Y.A.C. and W.S. Kim (2013) Desalination 329: 29-34
10. Jande, Y.A.C. and W.S. Kim (2013) Separation and Purification Technology 115: 224-230
11. Garcia - Quismondo, E., R. Gomez, F. Vaquero, A.L. Cudero, J. Palma, and M.A. Anderson (2013) Physical Chemistry Chemical Physics 15: 7648-7656
12. Wang, G., B. Qian, Q. Dong, J. Yang, Z. Zhao, and J. Qiu (2013) Separation and Purification Technology 103: 216-221
13. Oren, Y. and A. Soffer (1983) Journal of Applied Electrochemistry 13: 473-487
14. Levi, M.D., G. Salitra, N. Levy, D. Aurbach, and J. Maier (2009) Nature Materials 8: 872-875
15. Avraham, E., Y. Bouhadana, A. Soffer, and D. Aurbach (2009) Journal of the Electrochemical Society 156: 95-99
16. Zhao, R., P.M. Biesheuvel, H. Miedema, H. Bruning, and A. van der Wal (2010) Journal of Physical Chemistry Letters 1: 205-210
17. Levi, M.D., N. Levy, S. Sigalov, G. Salitra, D. Aurbach, and J. Maier (2010) Journal of the American Chemical Society 132: 13220-13222
18. Kastening, B. and M. Heins (2001) Physical Chemistry Chemical Physics 3: 372-373
19. Han, L., K.G. Karthikeyan, M.A. Anderson, K. Gregory, J.J. Wouters, and A. Abdel-Wahab (2013) Electrochimica Acta 90: 573-581
20. Mossad, M. and L. Zou (2013) Chemical Engineering Journal 223: 704-713
21. Zhao, R., P.M. Biesheuvel, and A. Van der Wal (2012) Energy & Environmental Science 5: 9520-9527
22. Wu, P., J. Huang, V. Meunier, B.G. Sumpter, and R. Qiao (2012) The Journal of Physical Chemistry Letters 3: 1732-1737
23. Bazant, M.Z., K. Thornton, and A. Ajdari (2004) Physical Review E: Statistical, Nonlinear, and Soft Matter Physics 70: 021506
24. Biesheuvel, P.M. and M.Z. Bazant (2010) Physical Review E: Statistical, Nonlinear, and Soft Matter Physics E81: 031502
25. Zhao, R., M. van Soestbergen, H.H.M. Rijnaarts, A. van der Wal, M.Z. Bazant, and P.M. Biesheuvel (2012) Journal of Colloid and Interface Science 384: 38-44
26. Yang, K.-L., T.-Y. Ying, S. Yiacoumi, C. Tsouris, and E.S. Vittoratos (2001) Langmuir 17: 1961-1969

27. Gabelich, C.J., T.D. Tran, and I.H. Suffet (2002) Environmental Science & Technology 36: 3010-3019
28. Hou, C.-H., C. Liang, S. Yiacoumi, S. Dai, and C. Tsouris (2006) Journal of Colloid and Interface Science 302: 54-61
29. Xu, P., J.E. Drewes, D. Heil, and G. Wang (2008) Water Research 42: 2605-2617
30. Li, L., L. Zou, H. Song, and G. Morris (2009) Carbon 47: 775-781
31. Gabelich, C.J., P. Xu, and Y. Cohen (2010) Sustainability Science and Engineering 2: 295-326
32. Tsouris, C., R. Mayes, J. Kiggans, K. Sharma, S. Yiacoumi, D. DePaoli, and S. Dai (2011) Environmental Science & Technology 45: 10243-10249
33. Porada, S., L. Weinstein, R. Dash, A. van der Wal, M. Bryjak, Y. Gogotsi, and P.M. Biesheuvel (2012) ACS Applied Materials & Interfaces 4: 1194-1199
34. Porada, S., L. Borchardt, M. Oschatz, M. Bryjak, J. Atchison, K.J. Keesman, S. Kaskel, M. Biesheuvel, and V. Presser (2013) Energy & Environmental Science 6: 3700-3712
35. Lin, C., J.A. Ritter, and B.N. Popov (1999) Journal of The Electrochemical Society 146: 3639-3643
36. Kim, T. and J. Yoon (2013) Journal of Electroanalytical Chemistry 704: 169-174
37. Sharma, K., R.T. Mayes, J.O. Kiggans Jr, S. Yiacoumi, J. Gabitto, D.W. DePaoli, S. Dai, and C. Tsouris (2013) Separation and Purification Technology 116: 206-213
38. Biesheuvel, P.M., R. Zhao, S. Porada, and A. van der Wal (2011) Journal of Colloid and Interface Science 360: 239-248
39. Porada, S., M. Bryjak, A. van der Wal, and P.M. Biesheuvel (2012) Electrochimica Acta 75: 148-156
40. Rica, R.A., D. Brogioli, R. Ziano, D. Salerno, and F. Mantegazza (2012) The Journal of Physical Chemistry C 116: 16934–16938
41. Porada, S., B.B. Sales, H.V.M. Hamelers, and P.M. Biesheuvel (2012) The Journal of Physical Chemistry Letters 3: 1613-1618
42. Andersen, M.B., M. van Soestbergen, A. Mani, H. Bruus, P.M. Biesheuvel, and M.Z. Bazant (2012) Physical Review Letters 109: 108301
43. Galama, A.H., J.W. Post, M.A. Cohen Stuart, and P.M. Biesheuvel (2013) Journal of Membrane Science 442: 131-139
44. Biesheuvel, P.M., W.M. de Vos, and V.M. Amoskov (2008) Macromolecules 41: 6254-6259
45. de Vos, W.M., P.M. Biesheuvel, A. de Keizer, J.M. Kleijn, and M.A. Cohen Stuart (2009) Langmuir 25: 9252-9261
46. Biesheuvel, P.M. (2004) Journal of Physics: Condensed Matter 16: L499-L504
47. Spruijt, E. and P.M. Biesheuvel (2014) Journal of Physics: Condensed Matter 26:
48. Huang, J., R. Qiao, G. Feng, B.G. Sumpter, and V. Meunier, *Modern Theories of Carbon-Based Electrochemical Capacitors*, in *Supercapacitors*. 2013, Wiley-VCH Verlag GmbH & Co. KGaA. p. 167-206.
49. Garten, V.A. and D.E. Weiss (1955) Australian Journal of Chemistry 8: 68-95
50. Garten, V.A. and D.E. Weiss (1957) Reviews of pure and applied chemistry 7: 69-122
51. Attard, P., D.J. Mitchell, and B.W. Ninham (1988) The Journal of Chemical Physics 89: 4358-4367
52. Skinner, B., M.S. Loth, and B.I. Shklovskii (2010) Physical Review Letters 104: 128302
53. Biesheuvel, P.M., Y.Q. Fu, and M.Z. Bazant (2011) Physical Review E: Statistical, Nonlinear, and Soft Matter Physics E83: 061507
54. Biesheuvel, P.M., Y. Fu, and M.Z. Bazant (2012) Russian Journal of Electrochemistry 48: 580-592
55. Rica, R.A., R. Ziano, D. Salerno, F. Mantegazza, M.Z. Bazant, and D. Brogioli (2013) Electrochimica Acta 92: 304– 314
56. Hou, C.-H., T.-S. Patricia, S. Yiacoumi, and C. Tsouris (2008) The Journal of Chemical Physics 129: 224703-224709
57. Feng, G., R. Qiao, J. Huang, B.G. Sumpter, and V. Meunier (2010) ACS Nano 4: 2382-2390
58. Bonthuis, D.J., S. Gekle, and R.R. Netz (2011) Physical Review Letters 107: 166102
59. Feng, G. and P.T. Cummings (2011) The Journal of Physical Chemistry Letters 2: 2859-2864
60. Kondrat, S. and A. Kornyshev (2011) Journal of Physics: Condensed Matter 23: 022201
61. Jadhao, V., F.J. Solis, and M.O. de la Cruz (2013) The Journal of Chemical Physics 138: 054119-13
62. Jiménez, M.L., M.M. Fernández, S. Ahualli, G. Iglesias, and A.V. Delgado (2013) Journal of Colloid and Interface Science 402: 340-349

63. Wang, H., A. Thiele, and L. Pilon (2013) The Journal of Physical Chemistry C 117: 18286-18297
64. Kobrak, M.N. (2013) Journal of Physics: Condensed Matter 25: 095006
65. Kastening, B. and M. Heins (2005) Electrochimica Acta 50: 2487-2498
66. Suss, M.E., T.F. Baumann, M.A. Worsley, K.A. Rose, T.F. Jaramillo, M. Stadermann, and J.G. Santiago (2013) Journal of Power Sources 241: 266–273
67. Grahame, D.C. (1947) Chemical Reviews 41: 441-501
68. Bazant, M.Z., K.T. Chu, and B.J. Bayly (2005) SIAM Journal on Applied Mathematics 65: 1463-1484
69. Kalluri, R.K., M.M. Biener, M.E. Suss, M.D. Merrill, M. Stadermann, J.G. Santiago, T.F. Baumann, J. Biener, and A. Striolo (2013) Physical Chemistry Chemical Physics 15: 2309-2320
70. Andersen, P.S. and M. Fuchs (1975) Biophysical journal 15: 795-830
71. Grosberg, A.Y., T.T. Nguyen, and B.I. Shklovskii (2002) Reviews of Modern Physics 74: 329-345
72. Santangelo, C.D. (2006) Physical Review E 73: 041512
73. Hatlo, M.M. and L. Lue (2010) EPL (Europhysics Letters) 89: 25002
74. Bazant, M.Z., B.D. Storey, and A.A. Kornyshev (2011) Physical Review Letters 106: 046102
75. Jackson, J.D., *Classical electrodynamics*. Second edition ed. 1975: Wiley. 848.
76. Zhao, R., O. Satpradit, H.H.M. Rijnaarts, P.M. Biesheuvel, and A. van der Wal (2013) Water Research 47: 1941-1952
77. Zhao, R., S. Porada, P.M. Biesheuvel, and A. van der Wal (2013) Desalination 330: 35-41
78. Levi, M.D., S. Sigalov, D. Aurbach, and L. Daikhin (2013) The Journal of Physical Chemistry C 117: 14876–14889
79. Müller, M. and B. Kastening (1994) Journal of Electroanalytical Chemistry 374: 149-158
80. Gupta, V.K., D. Pathania, S. Sharma, and P. Singh (2013) Journal of Colloid and Interface Science 401: 125-132
81. Aghakhani, A., S.F. Mousavi, B. Mostafazadeh-Fard, R. Rostamian, and M. Seraji (2011) Desalination 275: 217-223
82. Tarazona, P. (1985) Physical Review A 31: 2672-2679
83. van Soestbergen, M. (2013) Personal Communication:
84. Dlugolecki, P. and A. van der Wal (2013) Environmental Science & Technology 47: 4904–4910
85. Liang, P., L. Yuan, X. Yang, S. Zhou, and X. Huang (2013) Water Research 47: 2523-2530
86. Kim, Y.-J., J.-H. Kim, and J.-H. Choi (2013) Journal of Membrane Science 429: 52-57
87. Yeo, J.-H. and J.-H. Choi (2013) Desalination 320: 10-16
88. Paz-Garcia, J.M., O. Schaetzle, P.M. Biesheuvel, and H.V.M. Hamelers (2014) Journal of Colloid and Interface Science 418: 200–207
89. Hamelers, H.V.M., O. Schaetzle, J.M. Paz-García, P.M. Biesheuvel, and C.J.N. Buisman (2014) Environmental Science & Technology Letters 1: 31–35
90. Grochowski, P. and J. Trylska (2008) Biopolymers 89: 93-113
91. Levitt, D.G. (1986) Annual Review of Biophysics and Biophysical Chemistry 15: 29-57
92. Levi, M.D., S. Sigalov, G. Salitra, D. Aurbach, and J. Maier (2011) ChemPhysChem 12: 854-862
93. Levi, M.D., S. Sigalov, G. Salitra, R. Elazari, and D. Aurbach (2011) The Journal of Physical Chemistry Letters 2: 120-124